\documentclass[11pt]{article}

\usepackage[preprint]{acl}

\usepackage{times}
\usepackage{latexsym}

\usepackage[T1]{fontenc}

\usepackage[utf8]{inputenc}

\usepackage{microtype}

\usepackage{inconsolata}

\usepackage{graphicx}
\usepackage{booktabs}
\usepackage{amsmath} 

\title{Heterogeneous LLM Debate Under Adversarial Peers: Honest Gains, Replacement Costs, and Resilience}

\author{
Prashanti Nilayam \quad
Kiran Kumar Ramanna \quad
Prashil Tumbade \quad
Sankalp Nayak \\
ServiceNow\\
}

\begin{document}
\maketitle

\begin{abstract}
Heterogeneous LLM debate is motivated by the promise that diverse peers correct one another, but the same exchange that carries correction also carries adversarial influence. We measure which dominates by tracking how a heterogeneous peer changes the honest agents' revision behavior: how often they change their answer, and whether the change is corrective or harmful. We compare matched panels (homogeneous baseline, honest-mixed, and adversarial-mixed) and contaminated panels in which a malicious same-family peer is already present, spanning four model families and three reasoning benchmarks. An honest heterogeneous peer sharply lowers harmful revision, and an adversarial one reverses it. For \texttt{Llama-3.1-70B} defenders on MATH-hard, the honest-slot harmful-revision rate falls from $89\%$ in the homogeneous panel to $35\%$ with an honest peer, and an adversarial peer returns it to $90\%$. The conditional rate hides this damage on weak defenders, but the end-of-debate flip rate exposes it. The pattern keeps its sign across families and benchmarks while its magnitude varies with the defender--benchmark regime. We also measure the effects when an adversarial same-family peer is already present: an honest heterogeneous peer lowers both harmful revision and the rate at which initially-correct answers are lost. On the same \texttt{Llama-3.1-70B} setting, the added honest peer cuts the flip rate on initially-correct items from $31\%$ under a same-family adversary to $6\%$. Heterogeneity is therefore not only an attack surface but, when an adversary is already present, also a defense.
\end{abstract}
\section{Introduction}
\label{sec:intro}

Multi-agent debate improves language-model reasoning by having several models answer independently and revise through exchanged arguments \citep{du2024improving}, and heterogeneous panels are further motivated by evidence that model diversity contributes complementary strengths and decorrelated errors \citep{hegazy2024diversity, zhou2025adaptive}. Revision is double-edged: models often revise correct answers into wrong ones \citep{huang2024large}, and an adversarial or systematically misleading peer can exploit the same exchange that supports honest correction \citep{amayuelas2024multiagent}. A heterogeneous peer is therefore as much a channel for influence as for evidence.

We study what heterogeneity is worth once that channel can be attacked. Rather than scoring panels by final accuracy, we measure how a heterogeneous peer changes the \emph{revision behavior of the honest agents}: how often they change their answer, and whether the change is corrective or harmful. This defender-centered lens separates the benefit a peer adds from the harm it can do, in the two settings a deployer actually faces: \textbf{(i)} the added peer is itself the adversary and \textbf{(ii)} the panel is already contaminated, and we swap in an \emph{honest} heterogeneous peer. Can diversity defend a panel where an adversary is already present?

The two settings give one answer. An honest heterogeneous peer produces a large reduction in harmful revision, and a single adversarial peer erases it: for \texttt{Llama-3.1-70B} defenders on MATH-hard, the honest-slot harmful-revision rate moves $89\% \to 35\% \to 90\%$ across the homogeneous, honest-mixed, and adversarial-mixed panels. The adversary's cost is the destruction of a large latent gain. And when an adversary is already present, an honest heterogeneous peer acts as a defense, sharply reducing both harmful revision and end-of-debate regressions. Both effects keep the same sign across four model families and three benchmarks while their magnitude varies with the defender--benchmark regime.

\paragraph{Contributions.}
\begin{itemize}
\item A defender-centered measurement of debate that separates beneficial from harmful revision, and the masking effect it exposes: on weak defenders the harmful-revision rate saturates near its ceiling, while the flip rate on initially-correct items reveals damage the conditional rate hides.
\item  A matched benefit--risk tradeoff: the honest bonus, the adversarial penalty, and the replacement cost of swapping an honest peer for an adversarial one. The tradeoff's sign is invariant across families and benchmarks; its magnitude varies across defender--benchmark regimes.
\item A same-family adversary control showing that adversarial harm does not require a heterogeneous attacker. A same-family adversary degrades revision in the same regime and the effect is not an artifact of model diversity.
\item A contamination result showing that an honest heterogeneous peer adds robustness when an adversary is already present, reducing harmful revision and final regressions on both \texttt{gpt-4.1} and \texttt{Llama-3.1-70B}, and on SciBench.
\end{itemize}

\section{Framework}
\label{sec:framework}

A heterogeneous debate panel is an influence network, and influence runs in both directions. The same channel that lets an honest peer contribute complementary evidence and correct a mistake also lets a malicious peer push an honest agent toward a wrong answer. Whether heterogeneity helps or harms is therefore not fixed by diversity alone; it depends on who is added to the panel and whether that peer can be trusted, and it is settled in how the honest agents revise.

We call a panel member an \emph{adversarial peer} when its objective is to induce harmful revision in the honest agents rather than to answer the task correctly; \S\ref{sec:setup} gives the construction.

\subsection{Method and Measurement}

\paragraph{Method.} Our experimental method is multi-agent debate \citep{du2024improving}. Three agents answer a question independently and then revise their answers over several rounds. At each round, every agent receives the others' previous-round responses alongside the question and produces a new answer, which may agree with or differ from its prior response. We treat the final-round answer as the agent's response. We hold this protocol fixed and vary only panel composition: which models fill the slots, and whether one of them is adversarial. Full protocol settings are given in \S\ref{sec:setup}.

\paragraph{Measurement.} Final accuracy alone cannot answer our question, because it hides the mechanism we care about: the direction in which honest agents change their answers under peer influence. We measure that direction with the detection--generation decomposition of \citet{nilayam2026detection}. From one round to the next, an honest agent either keeps its answer or changes it, and a change is either toward the correct answer or away from it. This yields four regimes (Table~\ref{tab:regimes}): \textbf{(i)} the agent keeps a correct answer (\textbf{BOUNDARY}), \textbf{(ii)} keeps an incorrect answer (\textbf{IP}), \textbf{(iii)} changes an incorrect answer to the correct one (\textbf{DC}, a corrective revision), or \textbf{(iv)} changes its answer to a wrong one (\textbf{DM}, a harmful revision). For panel composition the decisive contrast is DC against DM: not whether agents revise, but whether revision carries them toward truth or away from it.

\begin{table*}[t]
\centering
\small
\begin{tabular}{lcccl}
\toprule
Regime & $D$ & $T(\phi)$ & $G$ & Interpretation \\
\midrule
BOUNDARY & 0 & 1 & --- & correct answer kept \\
IP       & 0 & 0 & --- & incorrect answer kept \\
DC       & 1 & 0 & correct   & revised to correct \\
DM       & 1 & 0 & incorrect & revised to incorrect \\
\bottomrule
\end{tabular}
\caption{The four revision regimes. $D$ is detection (whether the agent changes its answer), $T(\phi)$ indicates whether the agent's current answer is correct, and $G$ is the generated answer (defined only when $D{=}1$). Revisions away from a correct answer ($D{=}1$, $T(\phi){=}1$) are also counted as DM, since they end in a wrong answer.}
\label{tab:regimes}
\end{table*}

Two rates summarize a panel's revision behavior. Let $N_{\mathrm{valid}}$ be the number of valid honest-slot transitions, $N_{\mathrm{chg}}$ the number whose parsed answer changes, and $N_{\mathrm{DM}}$ the number of those changes that land in DM. Then
\[
P(D{=}1)=\frac{N_{\mathrm{chg}}}{N_{\mathrm{valid}}},
\qquad
P(\mathrm{DM}\mid D{=}1)=\frac{N_{\mathrm{DM}}}{N_{\mathrm{chg}}}.
\]
The revision rate $P(D{=}1)$ is how often honest agents change their answer; the harmful-revision rate $P(\mathrm{DM}\mid D{=}1)$ is the share of those changes that move away from the correct answer. An honest peer should lower the harmful-revision rate; an adversary should raise it.

\subsection{The Matched Comparison: Honest Bonus and Adversarial Penalty}

We compare three panels that differ in a single slot: a \textbf{homogeneous baseline}; an \textbf{honest-mixed} panel that replaces one slot with an honest heterogeneous peer; and an \textbf{adversarial-mixed} panel that replaces the same slot with an adversarial one. Write $p_{\text{base}}$, $p_{\text{hon}}$, and $p_{\text{adv}}$ for the harmful-revision rate $P(\mathrm{DM}\mid D{=}1)$ under each. All three are points on one axis, and the comparison is easiest to read that way (\emph{lower is better}):
\[
\resizebox{\columnwidth}{!}{$\displaystyle
\underbrace{p_{\text{hon}}}_{\text{honest-mixed}}
\xleftarrow{\ \text{honest bonus}\ }
\underbrace{p_{\text{base}}}_{\text{baseline}}
\xrightarrow{\ \text{adv.\ penalty}\ }
\underbrace{p_{\text{adv}}}_{\text{adv-mixed}}
$}
\]
An honest peer moves the rate left by the \emph{honest bonus} $p_{\text{base}}-p_{\text{hon}}$; an adversary moves it right by the \emph{adversarial penalty} $p_{\text{adv}}-p_{\text{base}}$; the distance between the two peers, $p_{\text{adv}}-p_{\text{hon}} = \text{bonus} + \text{penalty}$, is the \emph{replacement cost} of swapping an honest peer for an adversarial one.

The replacement cost is what a deployer pays when a trusted heterogeneous peer is compromised. It can be large even when the adversarial penalty alone is small: most of the loss is the forgone honest bonus.

A deployer adds a heterogeneous peer without knowing its intent. If the peer is adversarial with prior probability $q$, heterogeneity still lowers expected harmful revision whenever $q$ falls below the break-even prior $q^\star = \text{bonus}/(\text{bonus}+\text{penalty})$; we report $q^\star$ for every matched setting in Appendix Table~\ref{tab:appendix_breakeven}.

\subsection{The Contamination Comparison: Heterogeneity as a Defense}

The matched comparison asks whether to \emph{add} a heterogeneous peer to a clean panel. A second question matters just as much in practice: what to do when a panel is \emph{already} contaminated. We compare a clean homogeneous panel, a same-family adversarial panel, and a contaminated mixed panel that keeps the adversary but replaces one honest same-family defender with an honest heterogeneous peer, and ask whether that swap helps or harms the remaining honest agents.

In this regime the harmful-revision rate is no longer sufficient on its own. When defenders are weak, $P(\mathrm{DM}\mid D{=}1)$ saturates near its ceiling: almost every revision in the clean panel is already harmful, so the rate has little room left to register an adversary even when the adversary is plainly changing outcomes (\S\ref{sec:results}). We therefore introduce the \emph{flip rate}, an end-to-end measure that is not subject to this ceiling. Let $\mathcal{C}_0$ be the set of items on which all honest slots are correct at $R0$, and $\mathcal{C}_F \subseteq \mathcal{C}_0$ those on which they all remain correct at the final round. The flip rate is 
\[ \text{flip} = \frac{|\mathcal{C}_0| -|\mathcal{C}_F|}{|\mathcal{C}_0|}.\]

Where the harmful-revision rate measures per-step revision quality, the flip rate measures end-to-end retention: among items initially correct on every honest slot, the fraction on which at least one slot becomes incorrect by the final round. The two rates measure different things by construction: one conditional and first-step, the other unconditional and end-to-end. The contamination results turn on the gap between them.
\section{Experimental Setup}
\label{sec:setup}

We run two families of experiments. The \emph{matched tradeoff} compares a homogeneous baseline against honest-mixed and adversarial-mixed panels that each replace a single slot, isolating the benefit and cost of one added peer. The \emph{contamination control} begins from a panel that already contains a malicious same-family peer and asks whether replacing one honest same-family defender with an honest heterogeneous peer helps or harms the remaining honest agents.

\subsection{Benchmarks and Analysis Scope}

We work across a capability ladder spanning three reasoning benchmarks. MATH-hard (levels 4--5 of the MATH-500 evaluation set; \citealp{hendrycks2021measuring, lightman2024let}) is the primary benchmark: it carries a deep pool of plausible wrong answers and supports clean matched and contaminated cohorts. SciBench extends the controls to a distinct scientific-reasoning domain, and GSM8K anchors the low-capability end of the ladder on easier arithmetic. Per-benchmark item counts, level filters, and decoding budgets are given in Table~\ref{tab:benchmark_config} (Appendix~\ref{app:cohorts}).

We analyze the first revision step, $R0 \to R1$, where panel composition has its cleanest interpretation: every honest agent has seen the same problem, and the only new input is the first round of peer responses. Later rounds compound direct peer influence with path dependence from earlier revisions, so we use $R0 \to R1$ for the main comparisons and read later rounds as cascade diagnostics.

All reported quantities are aggregated over $K{=}3$ independent replicates per cohort to account for run-to-run LLM variability under nominally deterministic decoding. We pool honest-slot transitions across replicates for the transition-level rates; the item-level flip rate is computed over items on which all honest slots are correct at $R0$ (\S\ref{sec:framework}).

\subsection{Panels}

\paragraph{Matched cohorts.} The matched comparison spans a capability ladder of capable models---small open-weight, large open-weight, and frontier open- and closed-weight---across the three benchmarks, so the pattern is tested rather than read off a single family. On MATH-hard the defenders are \texttt{Llama-3.1-70B}: homogeneous \texttt{[ll, ll, ll]}, honest-mixed \texttt{[ll, ll, gpt-oss-120b]}, and adversarial-mixed \texttt{[ll, ll, gpt-oss-120b-adv]}. We run the analogous \texttt{gpt-4.1} trio on MATH-hard and a small-model trio on GSM8K: homogeneous \texttt{[llama-1b, llama-1b, llama-1b]}, honest-mixed \texttt{[llama-1b, llama-1b, gemma-2b]}, and adversarial-mixed \texttt{[llama-1b, llama-1b, gemma-2b-adv]}. Within each trio the honest-mixed and adversarial-mixed panels differ only in the objective assigned to the added peer, which isolates adversariality from heterogeneity.

\paragraph{Contamination controls.} Each contamination setting compares three panels: a clean homogeneous panel, a same-family adversarial panel that turns one same-family slot adversarial, and a contaminated mixed panel that keeps that adversary but replaces one honest same-family defender with an honest heterogeneous peer. We run these on MATH-hard for both \texttt{Llama-3.1-70B} (\texttt{[ll,ll,ll]}, \texttt{[ll,ll,ll-adv]}, \texttt{[ll,ll-adv,gpt-oss-120b]}) and \texttt{gpt-4.1} (\texttt{[g,g,g]}, \texttt{[g,g,g-adv]}, \texttt{[g,g-adv,gpt-oss-120b]}), and we replicate the full structure on SciBench with \texttt{gpt-4.1} defenders and a \texttt{gpt-oss-120b} peer. The same-family panel holds capability fixed: the adversary is the defenders' own model. Any effect it produces therefore cannot be attributed to a more capable peer; we draw on this in \S\ref{sec:results}.

\subsection{Debate Protocol and Labeling}

We use the multi-agent debate protocol of \citet{du2024improving}: three agents, five rounds, synchronized so that round $r{+}1$ begins only after every agent has completed round $r$ and can see the others' round-$r$ responses. Decoding temperature is $0$ (or the model's minimum). Completion budgets are sized to expected solution length (4096 tokens for the long derivations of MATH-hard and SciBench, 1024 for GSM8K) and are listed per benchmark in Table~\ref{tab:benchmark_config}.

We label each honest-slot transition with the regimes of Table~\ref{tab:regimes} and the flip rate of \S\ref{sec:framework}, both assigned from the parsed boxed answer at each round. Transitions whose answer is unparseable on either side are excluded from the rates (parse-failure rates in Appendix~\ref{app:parsing}). In adversarial panels we exclude the designated adversary slot from all honest-slot counts; in a contaminated mixed panel such as \texttt{[ll, ll-adv, gpt-oss-120b]}, the honest pool is then one same-family defender and the honest heterogeneous peer. Some robustness analyses further restrict to \emph{adversary-effective} items: those on which the adversary's $R0$ answer is non-null and unequal to gold. This isolates items where the adversary actually committed to a wrong answer at the start of the debate.

\subsection{Adversarial Peer Construction}

An adversarial peer occupies one slot and is controlled entirely through its initial prompt (no fine-tuning, no protocol change), modeling an attacker who has compromised a single debate participant. It is subject to the same per-round completion budget and output structure as the honest agents, so it cannot gain influence simply by generating more than they do. For each item we precompute a plausible wrong target by perturbing the gold answer, then instruct the peer to commit to that target, give brief plausible reasoning for it, and skip verification or hedging; the prompt elicits a private \texttt{<reasoning>} block followed by a \texttt{<confident\_wrong\_response>} that the other agents see. When no numeric target can be generated, the peer is instructed to choose any plausible answer other than gold. We include an adversarial cohort once it reliably emits a non-parse-failed wrong answer at $R0$: the \texttt{gpt-oss-120b} adversaries on MATH-hard and SciBench, the same-family \texttt{gpt-4.1} adversary, and the \texttt{gemma-2b} adversary on GSM8K all meet this bar.

\section{Results}
\label{sec:results}

Across every setting we test, the pattern is the same: an honest heterogeneous peer lowers harmful revision, an adversarial one raises it, and the magnitude varies by defender--benchmark regime.

\subsection{The Matched Tradeoff}

\begin{figure*}[t]
  \centering
  \includegraphics[width=0.92\textwidth]{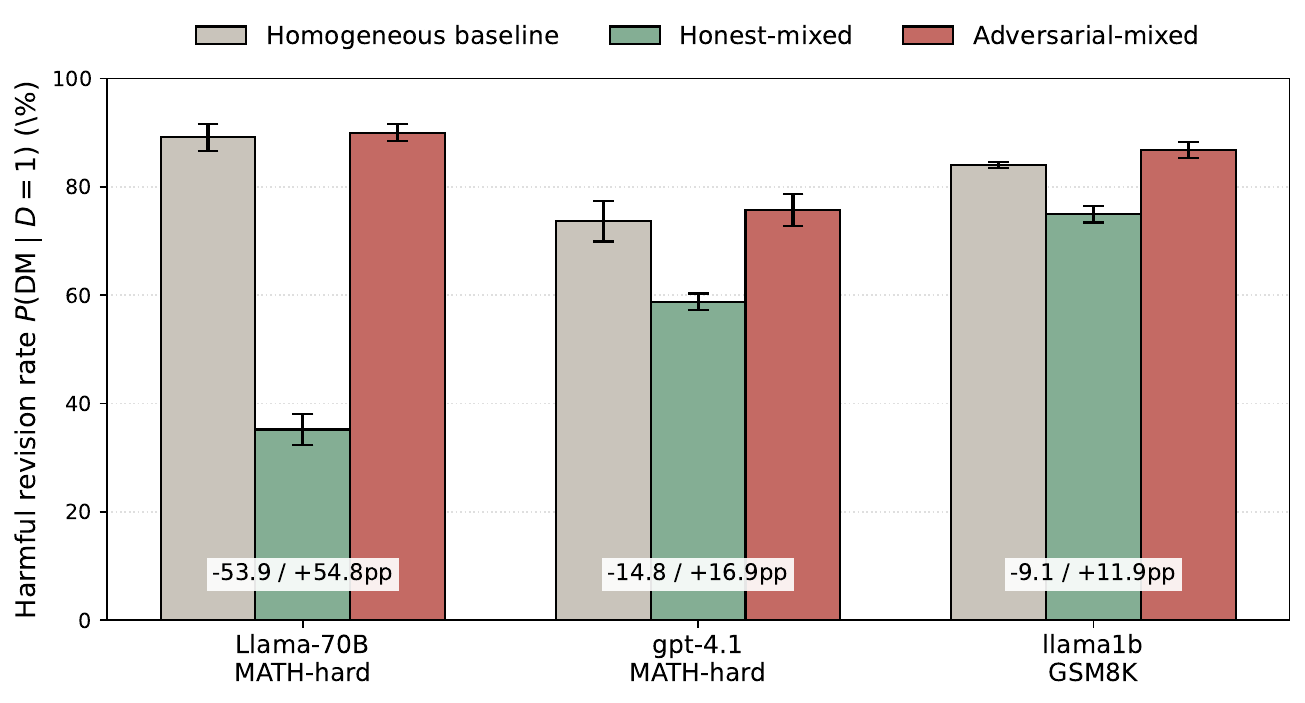}
  \caption{The matched tradeoff across settings. Each panel compares a homogeneous baseline, an honest heterogeneous peer, and an adversarial heterogeneous peer through the honest-slot harmful-revision rate $P(\mathrm{DM}\mid D{=}1)$ at the first revision step. The sign is the same across settings; the size of the honest bonus and replacement cost varies across defender--benchmark regimes.}
  \label{fig:tradeoff-multisetting}
\end{figure*}

The effect is largest for \texttt{Llama-3.1-70B} defenders on MATH-hard. In the homogeneous panel, honest defenders revise rarely ($P(D{=}1)=12.3\%\pm0.6$) and almost always harmfully ($p_{\text{base}}=89.1\%\pm2.5$). The honest heterogeneous peer more than triples the revision rate, to $42.2\%\pm0.4$, and cuts harmful revision to $p_{\text{hon}}=35.2\%\pm2.9$; the adversarial peer holds the revision rate high ($44.2\%\pm3.3$) but drives harmful revision back to $p_{\text{adv}}=90.0\%\pm1.5$. This is a quality shift, not volume: the honest peer triples revision, but the added revisions are overwhelmingly corrective---absolute corrective revision rises $1.3\% \to 27.3\%$ of transitions while harmful revision barely moves ($11.0\% \to 14.8\%$; Appendix Table~\ref{tab:appendix_matched_decomposition}). The honest bonus is $53.9$ points and the adversarial penalty $0.9$; the replacement cost of $54.8$ shows that the adversary's damage here is the destruction of a large honest gain. The gap reaches final accuracy: by the last round the honest panel holds $71\%$ honest-slot accuracy against the baseline's $42\%$, while the adversarial panel collapses to $22\%$ (Appendix Table~\ref{tab:appendix_matched_accuracy}).

The \texttt{gpt-4.1} trio on MATH-hard shows the same sign at smaller magnitude: harmful revision moves $73.6\% \to 58.8\% \to 75.7\%$ across baseline, honest-mixed, and adversarial-mixed (bonus $14.8$, penalty $2.1$). On GSM8K the small-model trio holds the sign with the smallest separation: $84.0\% \to 74.9\% \to 86.8\%$ (bonus $9.1$, penalty $2.8$). Across all three settings the break-even prior on adversarial peers is high (Appendix Table~\ref{tab:appendix_breakeven}): heterogeneity lowers expected harmful revision unless an added peer is adversarial with high probability. The sign is invariant; the magnitude is regime-dependent.

\subsection{Heterogeneity Restores Robustness Under Contamination}

\begin{figure*}[t]
  \centering
  \includegraphics[width=0.92\textwidth]{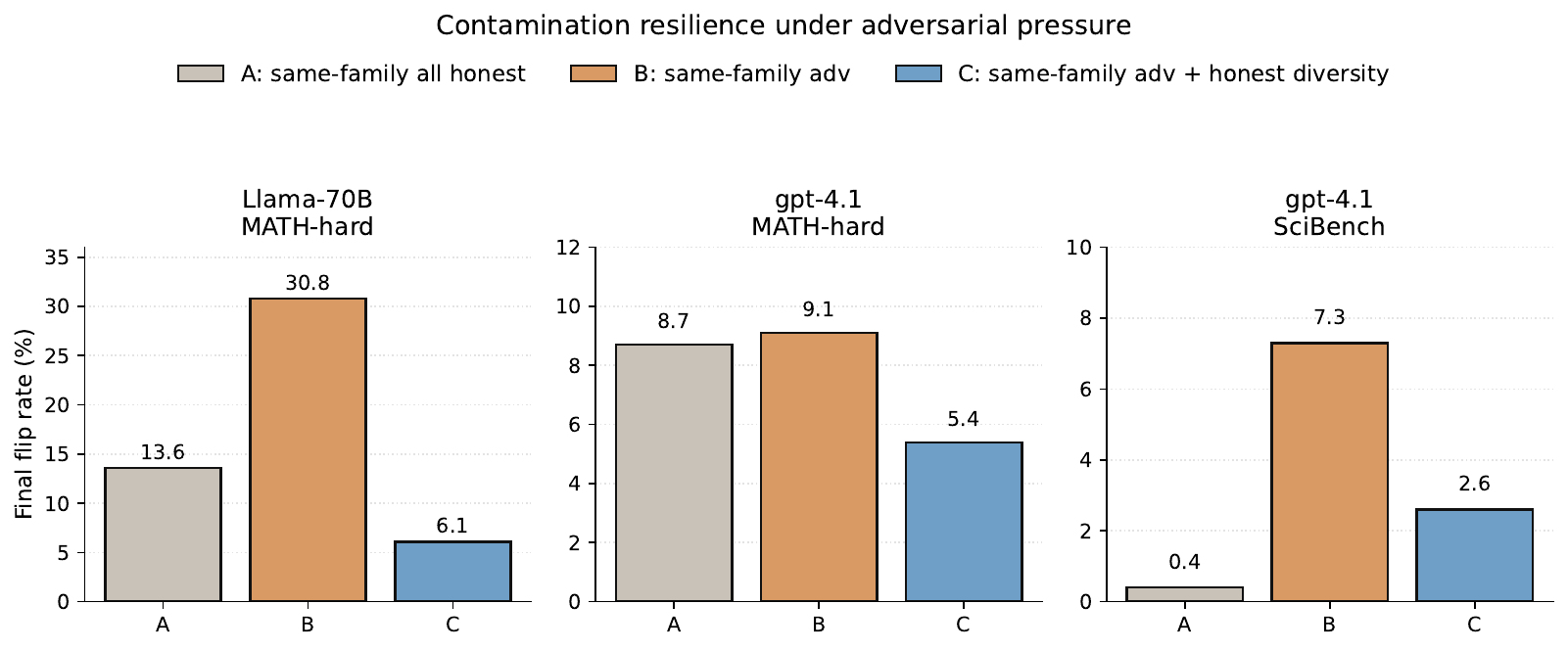}
  \caption{Contamination resilience under adversarial pressure. Each panel compares a same-family all-honest baseline, a same-family adversarial panel, and a contaminated mixed panel that keeps the same-family adversary while adding an honest heterogeneous peer. Bars report the final flip rate among items on which all honest slots are correct at $R0$. In every setting the contaminated mixed panel reduces final regressions relative to the same-family adversarial panel. Axis ranges differ across panels for readability.}
  \label{fig:contamination-flip}
\end{figure*}

When a panel already contains a malicious peer, replacing one honest same-family defender with an honest heterogeneous peer lowers both harmful revision and the flip rate in every setting and on both defender families (Table~\ref{tab:contamination-controls}, Figure~\ref{fig:contamination-flip}).

On MATH-hard with \texttt{Llama-3.1-70B}, the same-family adversarial panel barely moves harmful revision against the clean homogeneous panel ($87.2\%$ vs.\ $89.1\%$) yet doubles the flip rate on initially-correct items ($13.6\% \to 30.8\%$); the contaminated mixed panel cuts both ($43.7\%$ harmful revision, $6.1\%$ flip). A flat $P(\mathrm{DM}\mid D{=}1)$ beside a doubled flip rate is the ceiling effect of \S\ref{sec:framework}: the near-saturated conditional rate cannot register the adversary, while the end-to-end flip rate does. We trace the underlying item-level mechanism in \S\ref{sec:discussion}.

On \texttt{gpt-4.1}/MATH-hard the same-family adversarial panel tracks the clean homogeneous panel on both metrics (flip $9.1\%$ vs.\ $8.7\%$, harmful revision $72.5\%$ vs.\ $73.8\%$), while the contaminated mixed panel improves both (flip $5.4\%$, harmful revision $55.9\%$). On SciBench the same-family adversary drives the flip rate from $0.4\%$ to $7.3\%$ and harmful revision from $60.9\%$ to $64.4\%$, and the contaminated mixed panel pulls them back to $2.6\%$ and $56.4\%$. The defense holds across both benchmarks and both defender families.

\begin{table*}[t]
\centering
\small
\begin{tabular}{llcc}
\toprule
Setting & Panel & \shortstack{flip\%\\Wilson 95\% CI} & \shortstack{$P(\mathrm{DM}\mid D{=}1)$\\Wilson 95\% CI} \\
\midrule
\texttt{Llama-70B} / MATH-hard & clean \texttt{[ll,ll,ll]} & \shortstack{13.6\%\\$[10.4,17.6]$} & \shortstack{89.1\%\\$[84.8,92.3]$} \\
\texttt{Llama-70B} / MATH-hard & same-family \texttt{[ll,ll,ll-adv]} & \shortstack{30.8\%\\$[26.1,35.9]$} & \shortstack{87.2\%\\$[83.4,90.3]$} \\
\texttt{Llama-70B} / MATH-hard & contaminated \texttt{[ll,ll-adv,gpt-oss]} & \shortstack{6.1\%\\$[4.0,9.3]$} & \shortstack{43.7\%\\$[38.8,48.7]$} \\
\midrule
\texttt{gpt-4.1} / MATH-hard & clean \texttt{[g,g,g]} & \shortstack{8.7\%\\$[6.7,11.4]$} & \shortstack{73.8\%\\$[69.2,77.8]$} \\
\texttt{gpt-4.1} / MATH-hard & same-family \texttt{[g,g,g-adv]} & \shortstack{9.1\%\\$[7.0,11.8]$} & \shortstack{72.5\%\\$[66.9,77.4]$} \\
\texttt{gpt-4.1} / MATH-hard & contaminated \texttt{[g,g-adv,gpt-oss]} & \shortstack{5.4\%\\$[3.8,7.6]$} & \shortstack{55.9\%\\$[49.3,62.2]$} \\
\midrule
\texttt{gpt-4.1} / SciBench & clean \texttt{[g,g,g]} & \shortstack{0.4\%\\$[0.2,0.9]$} & \shortstack{60.9\%\\$[57.5,64.2]$} \\
\texttt{gpt-4.1} / SciBench & same-family \texttt{[g,g,g-adv]} & \shortstack{7.3\%\\$[6.0,8.8]$} & \shortstack{64.4\%\\$[60.8,67.7]$} \\
\texttt{gpt-4.1} / SciBench & contaminated \texttt{[g,g-adv,gpt-oss]} & \shortstack{2.6\%\\$[1.9,3.7]$} & \shortstack{56.4\%\\$[52.9,59.9]$} \\
\bottomrule
\end{tabular}
\caption{Contaminated-panel controls across defender families and benchmarks. The contaminated mixed panel keeps a malicious same-family peer but replaces one honest same-family defender with an honest heterogeneous peer. We report the final flip rate among items on which all honest slots are correct at $R0$, and the first-step harmful-revision rate $P(\mathrm{DM}\mid D{=}1)$. In every setting the contaminated mixed panel improves on the same-family adversarial control. Wilson intervals are item-level binomial for the flip rate and pooled-transition binomial for $P(\mathrm{DM}\mid D{=}1)$. Here \texttt{g} is \texttt{gpt-4.1}, \texttt{ll} is \texttt{Llama-3.1-70B}, and \texttt{gpt-oss} is \texttt{gpt-oss-120b}.}
\label{tab:contamination-controls}
\end{table*}

\section{Discussion}
\label{sec:discussion}

Honest heterogeneity helps through one specific mechanism rather than a generic aggregation effect: a heterogeneous peer shifts honest revisions from harmful to corrective, raising the corrective rate (DC) and lowering the harmful-revision rate (DM). The same channel that carries that correction carries adversarial influence, which is why a single malicious peer can erase the gain. Whether heterogeneity is worth adding therefore turns on defender capability, benchmark, and the integrity of the added peer, not on diversity alone.

The matched and contamination results expose two faces of this mechanism. In a clean panel, an honest heterogeneous peer improves revision quality and an adversarial one reverses it; what matters there is less the adversarial penalty than the replacement cost---the full gain destroyed when a trusted peer turns hostile. Under contamination the mechanism runs the other way: when a malicious peer is already present, adding an honest heterogeneous peer restores the remaining defenders' robustness, across both defender families and both benchmarks we test under contamination. Honest heterogeneity is not only a performance gain in benign panels; it is a defense in compromised ones.

Both standard intuitions about heterogeneous debate are correct, and they are the same story. Independent models contribute complementary evidence and reduce correlated failure; the same social influence also recruits honest agents into wrong revisions. These are not two phenomena but one revision channel, which (depending on the defenders and the benchmark) supports correction, amplifies harmful anchoring, or suppresses harmful drift.

The same-family \texttt{gpt-4.1} control on MATH-hard makes this concrete. There, the adversarial same-family and clean homogeneous panels finish with the same number of correct honest trajectories even though their item-level paths differ sharply (Appendix~\ref{app:item_paths}): the adversary anchors some debates on a plausible wrong answer and blocks recovery, but on others it suppresses the clean homogeneous panel's tendency to over-revise a correct answer into a wrong one. The two effects roughly cancel in aggregate accuracy and leave the near-saturated harmful-revision rate unmoved, while the flip rate, which counts only regressions from initially-correct items, still rises. This is why the contaminated mixed panel, not the same-family adversarial panel, is the revealing control: the honest heterogeneous peer does not merely undo an aggregate collapse; it breaks a regime in which harmful anchoring and harmful drift coexist.

The cross-setting differences carry as much information as the shared pattern. The same phenomenon appears across large open-weight, frontier closed, and small arithmetic-scale models, but its magnitude varies across defender--benchmark regimes. It is large for \texttt{Llama-3.1-70B} on MATH-hard, compressed for the small-model \texttt{llama-1b} on GSM8K. The sign is portable; the size is not. Across our settings the break-even prior stays high (Appendix Table~\ref{tab:appendix_breakeven}), so heterogeneity remains net-positive over a wide range of peer-integrity assumptions: a security-sensitive design parameter, set per deployment, rather than a default improvement.

\section{Related Work}
\label{sec:related}

\paragraph{Multi-agent debate and cross-model collaboration.}
Debate and cross-agent communication are widely used as test-time strategies for improving language-model reasoning. Early debate systems showed that multiple agents exchanging arguments improve factuality and reasoning on math and QA \citep{du2024improving}, and later work broadened the design space to divergent-thinking prompts and cross-model communication topologies \citep{liang2024encouraging, yin2023exchange}; others report that debate improves truthfulness when stronger or more persuasive models participate \citep{khan2024debating}. This line motivates heterogeneous panels as a source of complementary reasoning and treats peer interaction as cooperative by default. We measure what that interaction costs once a peer is adversarial.

\paragraph{Heterogeneity as useful diversity.}
A second line argues that heterogeneous panels outperform homogeneous ones because model diversity supplies complementary reasoning styles and decorrelated errors \citep{hegazy2024diversity, zhou2025adaptive}, extending a longer diversity-as-robustness lineage from ensemble methods \citep{yang2020dverge}. Prior evidence that same-architecture models share representational structure and failure modes \citep{patel2026representational} additionally motivates the same-family adversarial control in \S\ref{sec:setup}. We take those gains as given and quantify their exposure on the same panels: how much of the honest bonus a single adversarial peer can erase.

\paragraph{Conditional effectiveness and debate failure modes.}
Recent evaluations show debate is not uniformly beneficial: gains depend on model capability, task difficulty, and protocol \citep{yang2025revisiting}. Peer interaction can also suppress independent correction or converge a panel on wrong answers \citep{wynn2025talk, wu2025can, maryanskyy2026agents}. Adjacent work decomposes within-panel flips by distinct mechanisms \citep{hao2026not} or studies cooperative disagreement among benign agents \citep{wu2025hidden}; both differ from the compositional, adversarial framing we adopt. Our same-family adversary control isolates adversariality from heterogeneity by holding capability fixed (\S\ref{sec:setup}). We move the lens from whole-panel outcomes to the defender's revision step, and ask not only when debate fails but when an honest heterogeneous peer can repair it on a contaminated panel.

\paragraph{Adversarial influence in multi-agent systems.}
A growing body studies malicious or faulty participants in collaborative LLM systems, from resilience under faulty agents \citep{tse2408resilience} to adversarial influence inside debate and collaboration protocols \citep{kraidia2026collaboration, amayuelas2024multiagent, cui2025mad}; \citet{amayuelas2024multiagent} attacks debate via Best-of-N argument optimization and measures panel accuracy and agreement, while a separate line uses debate itself as a defense against adversarial inputs \citep{chern2024combating}. Related work on sycophancy in single-LLM settings \citep{ccelebi2025parrot} documents compliance dynamics at the model level rather than the panel level. Byzantine-fault-tolerant designs handle the same threat through weighted-trust or threshold consensus modifications \citep{jo2025byzantine, zheng2026rethinking}; we hold the standard debate protocol fixed and vary only composition. The closest measurement-style relative is CW-POR \citep{agarwal2025persuasion}, which reports judge-side persuasion override; our flip rate is defender-side, capturing whether honest agents revise away from a correct answer rather than whether a downstream judge selects the adversarial output. These works measure whole-system degradation, attack success, defense design, judge-side override, or protocol-level robustness. We measure the defender's revision behavior under matched and contaminated panels, which lets us separate the honest bonus, the adversarial penalty, the replacement cost, and the robustness an honest heterogeneous peer adds under contamination.

\section{Limitations}
\label{sec:limitations}

Our evidence spans multiple model families and three reasoning benchmarks, all with objective, checkable answers: the setting where revision direction can be measured unambiguously. We do not study open-ended generation or subjective tasks, where ``correct revision'' is not crisply defined. Whether the same revision dynamics carry over remains open.

Our threat model is one realistic attack rather than an exhaustive taxonomy. The adversary is a single prompt-level peer that commits to a plausible wrong answer---the compromised-participant case---and we do not model multi-peer collusion, adaptive optimization across items, weight-level modification, or attacks on the parser or evaluator. Across the four attack-sophistication variants we instantiate, we observe no consistent effect on flip rate or harmful revision (Appendix~\ref{app:attack_quality}); but because our prompt-level adversary produces transparently checkable errors on objective-answer tasks, genuinely deceptive attacks (and influence that compounds over later rounds) remain uncharacterized. The reported effects, including the break-even prior, are calibrated to this attack class and the first revision step at which we measure. Characterizing the broader attack surface is future work.

Our analysis is defender-centered and transition-level by design: it isolates the effect of peer composition on how honest agents revise. Operators may also care about final panel accuracy, calibration, or downstream decision quality, and we report the end-of-debate flip rate alongside the transition-level rates (Table~\ref{tab:contamination-controls}), with round-by-round accuracy and item-level paths in the appendix (Table~\ref{tab:appendix_matched_accuracy}, Appendix~\ref{app:item_paths}). Defender-centered revision metrics are the right tool for the composition question, and complementary to system-level evaluation.

\section{Ethics Statement}
\label{sec:ethics}

This work studies how a malicious peer influences honest agents in collaborative LLM systems, and the dual-use risk is direct: the same measurements could inform more effective attacks on deployed multi-agent workflows. We keep the emphasis defensive and diagnostic. Our goal is to quantify when heterogeneity is worth the surface it creates. We describe the adversary only at the level needed to reproduce our measurements, not as an optimized or transferable attack.

The threat we model, a single compromised debate participant, is realistic for pipelines that incorporate third-party or tool-provided model outputs, and documenting its effect matters precisely because multi-agent debate is increasingly proposed as a reliability mechanism. Characterizing these failure modes supports more cautious deployment; it does not imply that heterogeneous panels are categorically unsafe or categorically beneficial.

Our experiments use standard academic benchmarks and involve no human subjects or sensitive personal data.

\bibliography{custom}

\appendix

\section{Cohort and Run Details}
\label{app:cohorts}

This appendix records the cohort-level details underlying the matched and contaminated-panel comparisons. Unless otherwise noted, all primary results are computed from $K=3$ independent replicates and focus on the first debate transition $R0 \to R1$. Adversarial analyses exclude the designated adversary slot from honest-slot regime counts.

\paragraph{Primary matched cohorts.}
Our main matched-tradeoff results come from three settings:
\begin{itemize}
\item \textbf{Llama-70B on MATH-hard:} homogeneous baseline \texttt{[llama-70b, llama-70b, llama-70b]}, honest-mixed \texttt{[llama-70b, llama-70b, gpt-oss-120b]}, and adversarial-mixed \texttt{[llama-70b, llama-70b, gpt-oss-120b-adv]}.
\item \textbf{\texttt{gpt-4.1} on MATH-hard:} homogeneous baseline \texttt{[gpt-4.1, gpt-4.1, gpt-4.1]}, honest-mixed \texttt{[gpt-4.1, gpt-4.1, gpt-oss-120b]}, and adversarial-mixed \texttt{[gpt-4.1, gpt-4.1, gpt-oss-120b-adv]}.
\item \textbf{small-model GSM8K:} homogeneous baseline \texttt{[llama-1b, llama-1b, llama-1b]}, honest-mixed \texttt{[llama-1b, llama-1b, gemma-2b]}, and adversarial-mixed \texttt{[llama-1b, llama-1b, gemma-2b-adv]}.
\end{itemize}

\paragraph{Primary contaminated-panel controls.}
The contamination controls span two defender families across three settings:
\begin{itemize}
\item \textbf{Llama-70B on MATH-hard:} clean \texttt{[llama-70b, llama-70b, llama-70b]}, same-family adversarial \texttt{[llama-70b, llama-70b, llama-70b-adv]}, and contaminated mixed \texttt{[llama-70b, llama-70b-adv, gpt-oss-120b]}.
\item \textbf{\texttt{gpt-4.1} on MATH-hard:} clean \texttt{[gpt-4.1, gpt-4.1, gpt-4.1]}, same-family adversarial \texttt{[gpt-4.1, gpt-4.1, gpt-4.1-adv]}, and contaminated mixed \texttt{[gpt-4.1, gpt-4.1-adv, gpt-oss-120b]}.
\item \textbf{\texttt{gpt-4.1} on SciBench:} clean \texttt{[gpt-4.1, gpt-4.1, gpt-4.1]}, same-family adversarial \texttt{[gpt-4.1, gpt-4.1, gpt-4.1-adv]}, and contaminated mixed \texttt{[gpt-4.1, gpt-4.1-adv, gpt-oss-120b]}.
\end{itemize}

\paragraph{Benchmark configuration.}
Table~\ref{tab:benchmark_config} gives the per-benchmark run configuration. All benchmarks use the same three-agent, five-round protocol at temperature $0$; completion budgets are sized to expected solution length (longer derivations on MATH-hard and SciBench, short answers on GSM8K). Item counts are the evaluated set after filtering: MATH-hard is restricted to levels 4--5 of MATH-500, and SciBench and GSM8K are the evaluable subsets of the sampled pools (nominal pools 500, 1000, and 2000 respectively).

\begin{table*}[t]
\centering
\small
\begin{tabular}{lccccc}
\toprule
Benchmark & items & levels & rounds & agents & output token budget \\
\midrule
MATH-hard & 262 & 4--5 & 5 & 3 & 4096 \\
SciBench  & 682 & --- & 5 & 3 & 4096 \\
GSM8K     & 1319 & --- & 5 & 3 & 1024 \\
\bottomrule
\end{tabular}
\caption{Per-benchmark run configuration. Items are the evaluated set ($K{=}3$ replicates each); temperature is $0$ throughout; budget is the per-round completion-token limit. MATH-hard is levels 4--5 of MATH-500.}
\label{tab:benchmark_config}
\end{table*}

\section{Adversarial Peer Construction and Compliance}
\label{app:adversary}

\subsection{Prompt-level adversarial peers}

All adversarial cohorts use a single prompt-level adversarial peer rather than a fine-tuned or weight-modified attacker. For numerical-answer items, we precompute a plausible wrong target and prompt the adversary to commit to that target with brief supporting reasoning. When no perturbable target can be generated automatically, the adversary is instructed to choose a plausible answer different from gold. This setup intentionally tests whether an ordinary debate channel can be exploited by one compromised participant; it is not intended as an upper bound on the strongest possible attack.

\subsection{Adversary-effective items}

For adversarial cohorts, we sometimes restrict analysis to \emph{adversary-effective} items, defined as items on which the adversary's $R0$ parsed answer is both non-null and different from gold. The purpose of this filter is to ensure that the reported measurements describe honest-defender behavior under a true adversarial condition---a confidently wrong peer at the first revision opportunity---rather than under degraded experimental conditions where the adversary failed to commit. We do not use this filter to quantify how often an adversary fails to follow instructions; the failure rate is a methodological side effect of the experimental setup, not a measurement of interest.

Compliance rates on the adversarial cohorts differ by attacker family. The main \texttt{gpt-oss-120b} and same-family \texttt{gpt-4.1} adversarial cohorts are the cleanest. The Gemini-based attacker cohort on MATH-hard is methodologically weaker and is documented in \S\ref{app:adversary} rather than used for any primary claim.

\subsection{Gemini-attacker parse-failure analysis}

The Gemini-based attacker on MATH-hard exhibits a distinctive failure mode: a long ``verify loop'' that exhausts the token budget without ever producing a boxed answer. Across the three MATH-hard replicates for the cohort \texttt{[gpt-4.1, gpt-4.1, gemini-2.5-flash-adv]}, the adversary's $R0$ parse-fail rate is \textbf{189/786 = 24.0\%}. Of the 66 distinct items that parse-fail at least once, 61 fail in all three replicates. In every parse-fail event, the adversary hits the maximum token budget; about half the failures verbally commit to the target string somewhere in the output but never emit a boxed final answer, while the other half remain trapped in repeated re-verification.

These parse failures do not erase the peer-influence signal for the honest defenders. On the adversary-effective subset, the honest \texttt{gpt-4.1} defenders still show $P(D{=}1)=20.8\%\pm1.4$ and $P(\mathrm{DM}\mid D{=}1)=72.6\%\pm1.2$, close to the unfiltered numbers for the same cohort. We therefore treat the Gemini-attacker cohort as methodologically weaker than the cleaner attacker settings, but not as invalid.

\section{Extended Transition-Level Results}
\label{app:extended_results}

\subsection{Matched-tradeoff summaries}

Table~\ref{tab:appendix_matched_summary} collects our main matched-tradeoff summaries. These are the full numerical values behind Figure~\ref{fig:tradeoff-multisetting}.

\begin{table*}[t]
\centering
\small
\begin{tabular}{lcccc}
\toprule
Cohort & \shortstack{$P(D{=}1)$\\mean$\pm$SD} & \shortstack{$P(\mathrm{DM}\mid D{=}1)$\\mean$\pm$SD} & \shortstack{pooled\\$n(D{=}1)$} & \shortstack{pooled\\$n_{\mathrm{valid}}$} \\
\midrule
Homogeneous baseline \texttt{Llama-70B}$\times$3 (MATH-hard) & 12.3\% $\pm$ 0.6 & 89.1\% $\pm$ 2.5 & 175 & 1420 \\
Honest-mixed \texttt{Llama-70B/Llama-70B/gpt-oss} (MATH-hard) & 42.2\% $\pm$ 0.4 & 35.2\% $\pm$ 2.9 & 589 & 1397 \\
Adversarial-mixed \texttt{Llama-70B/Llama-70B/gpt-oss} (MATH-hard) & 44.2\% $\pm$ 3.3 & 90.0\% $\pm$ 1.5 & 589 & 1333 \\
\midrule
Homogeneous baseline \texttt{gpt-4.1}$\times$3 (MATH-hard) & 17.9\% $\pm$ 1.1 & 73.6\% $\pm$ 3.7 & 400 & 2236 \\
Honest-mixed \texttt{gpt-4.1/gpt-4.1/gpt-oss} (MATH-hard) & 20.6\% $\pm$ 1.5 & 58.8\% $\pm$ 1.5 & 310 & 1505 \\
Adversarial-mixed \texttt{gpt-4.1/gpt-4.1/gpt-oss} (MATH-hard) & 18.4\% $\pm$ 1.2 & 75.7\% $\pm$ 2.9 & 274 & 1486 \\
\midrule
Homogeneous baseline \texttt{llama-1b}$\times$3 (GSM8K) & 38.1\% $\pm$ 0.3 & 84.0\% $\pm$ 0.5 & 3018 & 7912 \\
Honest-mixed \texttt{llama-1b/llama-1b/gemma-2b} (GSM8K) & 42.9\% $\pm$ 0.7 & 74.9\% $\pm$ 1.5 & 3390 & 7908 \\
Adversarial-mixed \texttt{llama-1b/llama-1b/gemma-2b} (GSM8K) & 48.6\% $\pm$ 2.5 & 86.8\% $\pm$ 1.5 & 3833 & 7886 \\
\bottomrule
\end{tabular}
\caption{$R0 \to R1$ honest-slot transition summaries for the matched-tradeoff settings we emphasize in the main text. Here \texttt{gpt-oss} abbreviates \texttt{gpt-oss-120b}. All rows are computed from the saved trajectories using the same $R0 \to R1$ honest-slot transition logic as in the main text.}
\label{tab:appendix_matched_summary}
\end{table*}

Table~\ref{tab:appendix_breakeven} reports the derived honest bonus, adversarial penalty, replacement cost, and break-even prior $q^\star = \text{bonus}/(\text{bonus}+\text{penalty})$ for each matched setting. The break-even prior is high throughout: heterogeneity lowers expected harmful revision unless an added peer is adversarial with probability above $q^\star$.

\begin{table*}[t]
\centering
\small
\begin{tabular}{lcccc}
\toprule
Setting & bonus & penalty & repl.\ cost & $q^\star$ \\
\midrule
\texttt{Llama-70B} / MATH-hard & 53.9 & 0.9 & 54.8 & 98.4\% \\
\texttt{gpt-4.1} / MATH-hard & 14.8 & 2.1 & 16.9 & 87.6\% \\
small-model / GSM8K & 9.1 & 2.8 & 11.9 & 76.5\% \\
\bottomrule
\end{tabular}
\caption{Matched-tradeoff decision quantities (percentage points of $P(\mathrm{DM}\mid D{=}1)$). Honest bonus $=p_{\text{base}}-p_{\text{hon}}$, adversarial penalty $=p_{\text{adv}}-p_{\text{base}}$, replacement cost $=$ bonus $+$ penalty, and break-even prior $q^\star=\text{bonus}/(\text{bonus}+\text{penalty})$. $q^\star$ assumes symmetric utility between a gained correction and an induced error; under risk-averse deployment, where regressions are penalized more heavily than missed corrections, the operational threshold is strictly lower than $q^\star$.}
\label{tab:appendix_breakeven}
\end{table*}

Table~\ref{tab:appendix_matched_decomposition} decomposes the same matched-tradeoff settings into absolute corrective and harmful revision rates (computed from the summary table as $DC/n_{\mathrm{valid}} = n(D{=}1)(1-P(\mathrm{DM}\mid D{=}1))/n_{\mathrm{valid}}$ and the analogous expression for $DM$). The honest gain is a quality shift, not mere volume: for \texttt{Llama-70B} on MATH-hard the honest peer triples revision, but the added revisions are overwhelmingly corrective---absolute corrective revision rises from $1.3\%$ to $27.3\%$ of transitions while harmful revision barely moves ($11.0\%$ to $14.8\%$).

\begin{table*}[t]
\centering
\small
\begin{tabular}{lccc}
\toprule
Cohort & \shortstack{absolute\\$DC / n_{\mathrm{valid}}$} & \shortstack{absolute\\$DM / n_{\mathrm{valid}}$} & \shortstack{absolute\\$D{=}1 / n_{\mathrm{valid}}$} \\
\midrule
Homogeneous baseline \texttt{Llama-70B}$\times$3 (MATH-hard) & 1.3\% & 11.0\% & 12.3\% \\
Honest-mixed \texttt{Llama-70B/Llama-70B/gpt-oss} (MATH-hard) & 27.3\% & 14.8\% & 42.2\% \\
Adversarial-mixed \texttt{Llama-70B/Llama-70B/gpt-oss} (MATH-hard) & 4.4\% & 39.8\% & 44.2\% \\
\midrule
Homogeneous baseline \texttt{gpt-4.1}$\times$3 (MATH-hard) & 4.7\% & 13.2\% & 17.9\% \\
Honest-mixed \texttt{gpt-4.1/gpt-4.1/gpt-oss} (MATH-hard) & 8.5\% & 12.1\% & 20.6\% \\
Adversarial-mixed \texttt{gpt-4.1/gpt-4.1/gpt-oss} (MATH-hard) & 4.5\% & 14.0\% & 18.4\% \\
\midrule
Homogeneous baseline \texttt{llama-1b}$\times$3 (GSM8K) & 6.1\% & 32.0\% & 38.1\% \\
Honest-mixed \texttt{llama-1b/llama-1b/gemma-2b} (GSM8K) & 10.8\% & 32.1\% & 42.9\% \\
Adversarial-mixed \texttt{llama-1b/llama-1b/gemma-2b} (GSM8K) & 6.4\% & 42.2\% & 48.6\% \\
\bottomrule
\end{tabular}
\caption{Absolute decomposition of the matched-tradeoff settings at $R0 \to R1$. Unlike $P(\mathrm{DM}\mid D{=}1)$, these rates separate increased movement from improved revision quality by reporting corrective and harmful revisions directly as fractions of valid honest-slot transitions.}
\label{tab:appendix_matched_decomposition}
\end{table*}

Table~\ref{tab:appendix_matched_accuracy} reports honest-slot accuracy across the full debate trajectory for the same matched settings. In all three cases, the honest heterogeneous panel improves over the homogeneous baseline at $R1$ and retains that advantage through $R4$, although the final margin is much larger for stronger defenders than for the small-model GSM8K regime.

\begin{table*}[t]
\centering
\small
\begin{tabular}{lccccc}
\toprule
Cohort & $R0$ & $R1$ & $R2$ & $R3$ & $R4$ \\
\midrule
Homogeneous baseline \texttt{Llama-70B}$\times$3 (MATH-hard) & 44.3\% & 42.0\% & 41.1\% & 42.3\% & 41.8\% \\
Honest-mixed \texttt{Llama-70B/Llama-70B/gpt-oss} (MATH-hard) & 43.8\% & 69.8\% & 71.2\% & 71.1\% & 71.1\% \\
Adversarial-mixed \texttt{Llama-70B/Llama-70B/gpt-oss} (MATH-hard) & 43.7\% & 34.2\% & 24.4\% & 21.6\% & 21.8\% \\
\midrule
Homogeneous baseline \texttt{gpt-4.1}$\times$3 (MATH-hard) & 75.8\% & 75.6\% & 74.3\% & 74.7\% & 74.8\% \\
Honest-mixed \texttt{gpt-4.1/gpt-4.1/gpt-oss} (MATH-hard) & 75.9\% & 80.4\% & 80.9\% & 81.0\% & 81.4\% \\
Adversarial-mixed \texttt{gpt-4.1/gpt-4.1/gpt-oss} (MATH-hard) & 76.5\% & 75.3\% & 74.5\% & 74.4\% & 74.5\% \\
\midrule
Homogeneous baseline \texttt{llama-1b}$\times$3 (GSM8K) & 41.8\% & 40.8\% & 40.6\% & 40.4\% & 40.1\% \\
Honest-mixed \texttt{llama-1b/llama-1b/gemma-2b} (GSM8K) & 42.5\% & 46.1\% & 44.2\% & 43.9\% & 43.7\% \\
Adversarial-mixed \texttt{llama-1b/llama-1b/gemma-2b} (GSM8K) & 42.2\% & 36.2\% & 32.7\% & 33.5\% & 33.2\% \\
\bottomrule
\end{tabular}
\caption{Honest-slot accuracy across rounds for the matched-tradeoff settings. Values are pooled across the replicates. The honest heterogeneous panel improves over the homogeneous baseline at the first revision opportunity and retains that advantage through the final round in all three settings, with the strongest persistence on \texttt{Llama-70B} and the weakest on the small-model GSM8K regime.}
\label{tab:appendix_matched_accuracy}
\end{table*}

\subsection{Contaminated-panel summaries}

In the main text, we report a compact contamination table with flip rate and $P(\mathrm{DM}\mid D{=}1)$, and use Figure~\ref{fig:contamination-flip} to visualize the flip-rate pattern. Table~\ref{tab:appendix_contamination_summary} expands those summaries with $R1$ accuracy and $P(D{=}1)$ so that the appendix has a self-contained record of the main robustness analyses. Because the contamination summaries use different honest-slot conditioning and item pools than the matched-tradeoff summaries, identical panel names can legitimately have different marginal revision rates across the two analyses.

\begin{table*}[t]
\centering
\small
\resizebox{\textwidth}{!}{%
\begin{tabular}{llcccc}
\toprule
Setting & Panel & \shortstack{flip\%\\Wilson 95\% CI} & \shortstack{$R1$ accuracy\\Wilson 95\% CI} & \shortstack{$P(D{=}1)$\\Wilson 95\% CI} & \shortstack{$P(\mathrm{DM}\mid D{=}1)$\\Wilson 95\% CI} \\
\midrule
\texttt{Llama-70B} / MATH-hard & clean \texttt{[ll,ll,ll]} & \shortstack{13.6\%\\$[10.4,17.6]$} & \shortstack{42.1\%\\$[40.1,44.1]$} & \shortstack{21.8\%\\$[19.6,24.2]$} & \shortstack{89.1\%\\$[84.8,92.3]$} \\
\texttt{Llama-70B} / MATH-hard & same-family \texttt{[ll,ll,ll-adv]} & \shortstack{30.8\%\\$[26.1,35.9]$} & \shortstack{40.6\%\\$[38.2,43.0]$} & \shortstack{38.5\%\\$[35.5,41.7]$} & \shortstack{87.2\%\\$[83.4,90.3]$} \\
\texttt{Llama-70B} / MATH-hard & contaminated \texttt{[ll,ll-adv,gpt-oss]} & \shortstack{6.1\%\\$[4.0,9.3]$} & \shortstack{76.3\%\\$[74.1,78.3]$} & \shortstack{29.1\%\\$[26.7,31.6]$} & \shortstack{43.7\%\\$[38.8,48.7]$} \\
\midrule
\texttt{gpt-4.1} / MATH-hard & clean \texttt{[g,g,g]} & \shortstack{8.7\%\\$[6.7,11.4]$} & \shortstack{75.5\%\\$[73.8,77.2]$} & \shortstack{19.7\%\\$[18.0,21.5]$} & \shortstack{73.8\%\\$[69.2,77.8]$} \\
\texttt{gpt-4.1} / MATH-hard & same-family \texttt{[g,g,g-adv]} & \shortstack{9.1\%\\$[7.0,11.8]$} & \shortstack{74.8\%\\$[72.6,76.9]$} & \shortstack{20.3\%\\$[18.3,22.6]$} & \shortstack{72.5\%\\$[66.9,77.4]$} \\
\texttt{gpt-4.1} / MATH-hard & contaminated \texttt{[g,g-adv,gpt-oss]} & \shortstack{5.4\%\\$[3.8,7.6]$} & \shortstack{80.7\%\\$[78.7,82.6]$} & \shortstack{16.1\%\\$[14.3,18.1]$} & \shortstack{55.9\%\\$[49.3,62.2]$} \\
\midrule
\texttt{gpt-4.1} / SciBench & clean \texttt{[g,g,g]} & \shortstack{0.4\%\\$[0.2,0.9]$} & \shortstack{70.3\%\\$[69.1,71.4]$} & \shortstack{17.1\%\\$[16.0,18.2]$} & \shortstack{60.9\%\\$[57.5,64.2]$} \\
\texttt{gpt-4.1} / SciBench & same-family \texttt{[g,g,g-adv]} & \shortstack{7.3\%\\$[6.0,8.8]$} & \shortstack{69.1\%\\$[67.7,70.5]$} & \shortstack{22.5\%\\$[21.1,24.0]$} & \shortstack{64.4\%\\$[60.8,67.7]$} \\
\texttt{gpt-4.1} / SciBench & contaminated \texttt{[g,g-adv,gpt-oss]} & \shortstack{2.6\%\\$[1.9,3.7]$} & \shortstack{70.2\%\\$[68.8,71.6]$} & \shortstack{23.8\%\\$[22.4,25.3]$} & \shortstack{56.4\%\\$[52.9,59.9]$} \\
\bottomrule
\end{tabular}%
}
\caption{Expanded contaminated-panel summaries. The main text emphasizes flip rate and $P(\mathrm{DM}\mid D{=}1)$ because together they capture both end-of-debate regression on initially correct items and first-step harmful revision quality. We also report $R1$ accuracy and $P(D{=}1)$ for completeness. Here \texttt{g} abbreviates \texttt{gpt-4.1}, \texttt{ll} abbreviates \texttt{Llama-3.1-70B}, and \texttt{gpt-oss} abbreviates \texttt{gpt-oss-120b}.}
\label{tab:appendix_contamination_summary}
\end{table*}

\section{Item-Level Path Analysis for the Same-Family \texttt{gpt-4.1} Adversary}
\label{app:item_paths}

The MATH-hard same-family control is unusual because it washes out in aggregate. At $R4$, the clean homogeneous panel and the same-family adversarial panel finish with the same number of correct honest-slot trajectories (1173 each), even though their item-level paths differ. Table~\ref{tab:appendix_flip_patterns} summarizes the dominant flip categories across the shared honest-slot trajectories from all three replicas.

\begin{table}[t]
\centering
\small
\begin{tabular}{lp{5.3cm}}
\toprule
Count & Pattern \\
\midrule
78 & Clean finishes correct, same-family adversary finishes incorrect \\
78 & Same-family adversary finishes correct, clean finishes incorrect \\
27 & Both start correct; adversarial run corrupts a correct answer late \\
28 & Both start correct; clean run corrupts a correct answer late \\
16 & Both start wrong; clean recovers but adversarial run does not \\
13 & Both start wrong; adversarial run recovers but clean run does not \\
\bottomrule
\end{tabular}
\caption{Dominant flip categories for the MATH-hard same-family \texttt{gpt-4.1} adversary control. The near-symmetry of the first two rows explains why the aggregate control washes out despite substantial item-level differences.}
\label{tab:appendix_flip_patterns}
\end{table}

The item-level examples reflect two opposing mechanisms. On some items, the adversarial panel anchors the debate to a plausible wrong answer and prevents recovery; for example, a clean run that moves from an initially wrong radical expression to the correct integer answer may instead stay locked to the distractor under the adversarial run. On other items, the adversarial panel suppresses harmful drift, preserving an initially correct symbolic or exact answer on problems where the clean homogeneous panel over-revises into a nearby wrong numeric form. This symmetry is why the mixed contaminated panel is the more informative control: the honest heterogeneous peer does not merely reverse an obvious aggregate collapse, but instead breaks a revision regime in which harmful anchoring and harmful overthinking coexist.

\section{Attack-Quality Stress Test}
\label{app:attack_quality}

To test whether the effect depends on the specific adversarial prompt, we vary the quality of the adversary's message while holding the panel fixed at \texttt{[gpt-4.1, gpt-4.1, gpt-4.1-adv]} on MATH-hard ($n=100$ items per level): L0, a bare wrong answer with no reasoning; L1, a confident assertion with no rationale; L2, the brief plausible rationale used in the main cohorts; and L3, an elaborate fabricated derivation.

Within this controlled comparison, varying attack sophistication across the four prompt variants did not produce a consistent effect on flip rate or harmful revision. Across the four levels the flip rate ($7.5$, $5.4$, $5.8$, $10.0\%$) and the harmful-revision rate ($75.0$, $83.9$, $68.6$, $67.6\%$) show no monotonic trend with attack sophistication and stay within the range of the main same-family \texttt{gpt-4.1}/MATH cell (flip $9.1\%$, $P(\mathrm{DM}\mid D{=}1)=72.5\%$). The effect thus appears driven by the presence of a committed wrong peer rather than the quality of its argument, indicating the reported results are not an artifact of the specific adversarial prompt. We do not claim this generalizes to genuinely deceptive attacks: on objective-answer math the auto-generated adversary produces transparently checkable errors, so attack sophistication cannot be varied beyond this range here.

\section{Parsing and Evaluation Details}
\label{app:parsing}

\subsection{Answer extraction and parse-failure rates}

All transition labels are assigned from parsed final answers rather than from free-form reasoning traces. Elevated parse-failure can distort both the measured revision rate and the conditional DM rate, so the MATH cohorts use a 4096-token completion budget to keep adversary parse-failure low.

The strongest parse-failure concern in the current evidence base remains the Gemini-based adversarial cohort discussed above. By comparison, the main \texttt{gpt-oss-120b} adversarial cohort, the same-family \texttt{gpt-4.1} adversarial cohort, and the small-model GSM8K trio are materially cleaner. In the GSM8K small-model trio, for example, pooled parse-failure rates are negligible (0.02\% for homogeneous, 0.08\% for honest-mixed, and 0.35\% for adversarial-mixed), so the smaller effect size there should be interpreted as a genuine capability-scale difference rather than as a parsing artifact.

\subsection{Implementation notes}

The experimental pipeline logs per-round model outputs, parsed answers, gold answers, and token counts for each agent and replicate. Adversarial prompts, answer normalization, and transition labeling are all deterministic conditional on the underlying model outputs. This makes it possible to recompute the reported transition summaries from saved trajectories and stratify analyses by adversary compliance.

\end{document}